\documentclass[sigconf]{acmart}
\usepackage{booktabs}
\usepackage{comment}
\definecolor{awesome}{RGB}{70,130,180}
\makeatletter
\newsavebox{\acmbox@box}
\newlength{\acmbox@innerwd}
\newenvironment{acmbox}[1]{%
  \def\acmbox@title{#1}%
  \par\medskip\noindent
  \setlength{\fboxsep}{6pt}%
  \setlength{\fboxrule}{0.8pt}%
  \setlength{\acmbox@innerwd}{\dimexpr\linewidth-2\fboxsep-2\fboxrule\relax}%
  \begin{lrbox}{\acmbox@box}%
  \begin{minipage}{\acmbox@innerwd}%
  \setlength{\parindent}{0pt}%
  \setlength{\parskip}{0.35\baselineskip}%
  {\color{awesome}\bfseries \acmbox@title}\par\vspace{2pt}%
}{%
  \end{minipage}%
  \end{lrbox}%
  \noindent\fcolorbox{awesome}{gray!5}{\usebox{\acmbox@box}}%
  \par\medskip
}
\makeatother

\AtBeginDocument{%
  }

\setcopyright{acmlicensed}
\copyrightyear{2026}
\acmYear{2026}
\acmDOI{XXXXXXX.XXXXXXX}
\acmConference[EASE 2026]{The 30th International Conference on Evaluation and Assessment in Software Engineering}{Tue 9 - Fri 12 June 2026}{Glasgow, United Kingdom}
\acmISBN{978-1-4503-XXXX-X/2018/06}
\begin{document}

\title{A Longitudinal Analysis of Good First Issue Practices and Newcomer Pull Requests in Popular OSS Projects}

\author{Hirotatsu Hoshikawa}
\affiliation{%
  \institution{Nara Institute of Science and Technology}
  \country{Japan}
}
\email{hoshikawa.hirotatsu.hh4@naist.ac.jp}

\author{Hidetake Tanaka}
\affiliation{%
  \institution{Nara Institute of Science and Technology}
  \country{Japan}
}
\email{tanaka.hidetake.te0@naist.ac.jp}

\author{Kazumasa Shimari}
\affiliation{%
  \institution{Wakayama University}
  \country{Japan}
}
\email{shimari@wakayama-u.ac.jp}

\author{Raula Gaikovina Kula}
\affiliation{%
  \institution{The University of Osaka}
  \country{Japan}
}
\email{raula-k@ist.osaka-u.ac.jp}

\author{Kenichi Matsumoto}
\affiliation{%
  \institution{Nara Institute of Science and Technology}
  \country{Japan}
}
\email{matumoto@is.naist.jp}

\begin{abstract}
Open-source software (OSS) projects rely on effective newcomer onboarding to sustain their communities. OSS projects widely adopt ``good first issue'' (GFI) labels to highlight beginner-friendly tasks.
As development practices continue to evolve, understanding how these onboarding mechanisms change over time is important for both maintainers and researchers.
This study analyzes 406,826 issues and 1,117 newcomer GFI PRs across 37 popular GitHub repositories (30 of which use GFI labels) over a four-year period from July 2021 to June 2025. We find that while the proportion of issues with GFI labels remained stable during the first three years, it underwent a statistically significant decline beginning in January 2024, with substantial variation across projects not explained by repository age or programming language. Despite this supply-side decline, newcomer engagement with GFI issues remains stable at approximately 27\%, suggesting that GFI labels maintain consistent attractiveness. Examining the outcomes of this engagement, we find that the merge rate of newcomer GFI PRs declined from 61.9\% to 42.2\%. Initial PR characteristics such as description length and code size show no significant association with merge outcomes, indicating that success is not predicted by the quantitative characteristics of the initial submission alone. Together, these findings reveal a widening gap between stable newcomer interest in GFIs and the declining availability and success of GFI-based onboarding, underscoring the need for maintainers to sustain both GFI labeling and review support.
\end{abstract}



\maketitle

\section{Introduction}

The sustained development of open-source software (OSS) projects depends critically on the influx and retention of new contributors~\cite{10.1145/2675133.2675215,sholler2019ten}. However, newcomers often face significant technical and social barriers when attempting to make their first contribution, including difficulty finding suitable tasks, understanding complex codebases, and navigating unfamiliar development processes~\cite{6943482,10.1145/2675133.2675215}.

To address these challenges, OSS projects have widely adopted the practice of labeling certain issues as ``good first issue'' (GFI) to identify tasks suitable for beginners~\cite{10.1145/3368089.3409746}. Prior research has examined the usage patterns of GFI labels and proposed automated recommendation methods~\cite{10.1145/3510003.3510196,10.1145/3475716.3475789}. However, Tan et al.~\cite{10.1145/3368089.3409746} showed that many GFIs are not resolved by newcomers, indicating that challenges remain regarding the effectiveness of GFI labeling practices.

In recent years, the emergence of generative AI tools and Large Language Models (LLMs) has begun reshaping how developers write and review code, with potential implications for newcomer contribution quality and maintainer review practices. Understanding how GFI practices and newcomer behavior have evolved over time is important for maintaining healthy OSS communities.

However, how GFI labeling practices and newcomer GFI pull request outcomes have evolved longitudinally across mainstream popular OSS projects has not been examined, particularly as generative AI tools have become prevalent in software development.

In this study, we analyze four years (July 2021 to June 2025) of GFI practices across 37 popular OSS projects on GitHub. We address the following research questions:

\textbf{RQ1: How have GFI practices and newcomer engagement changed over the four-year period?}

We examine the trends in GFI ratio and newcomer engagement rates, investigating how these patterns have evolved over time.

\textbf{RQ2: How have the characteristics and task types of newcomer GFI PRs changed over time, and are these factors associated with merge rates?}

We classify GFI issues into task types (Bug, Feature, Documentation, Other) based on their labels and analyze pull requests addressing GFI-labeled issues (referred to as GFI PRs) submitted by newcomers, examining how merge rates differ by task type and over time. We also examine PR-level characteristics associated with merge outcomes.

We analyzed 406,826 issues, identifying 3,300 GFI-labeled issues and 1,117 newcomer GFI PRs. The GFI ratio declined significantly beginning in January 2024 (Pettitt test, $p<0.001$), with substantial variation across projects, while newcomer engagement remained stable at approximately 27\%. The merge rate of newcomer GFI PRs also declined, suggesting that the challenges extend beyond GFI availability to onboarding outcomes.

\section{Related Work}
\label{sec:related_work}
Attracting and retaining newcomers is critical for OSS sustainability, yet newcomers face well-documented social and technical barriers when joining projects~\cite{6943482,10.1145/2675133.2675215}. To address these barriers, GitHub introduced the Good First Issue (GFI) label to guide newcomers toward suitable tasks. Tan et al.~\cite{10.1145/3368089.3409746} provided a systematic analysis of GFI adoption, showing that many GFIs remain unresolved by newcomers. Subsequent work has proposed automated approaches to identify and recommend GFIs~\cite{10.1145/3510003.3510196,10.1145/3475716.3475789,10.1109/ASE56229.2023.00158}. Studies have also examined how newcomers and existing contributors differ in their task-selection strategies~\cite{10.1145/3544902.3546236}.

However, task availability and recommendation alone have proven insufficient. Tan et al.~\cite{10.1109/ICSE48619.2023.00064} demonstrated that direct mentoring support is the key driver of newcomer success, a finding echoed by work examining support from the maintainer side~\cite{Steinmacher2021,10.1145/3412569.3412571,10.1145/3510458.3513020,10.1145/2884781.2884806}. Studies of newcomer behavior further show that first contributions tend toward bug fixes~\cite{9273270} and that effective onboarding strategies vary by project~\cite{10.1109/TSE.2025.3550881}. While prior work has established the value of GFI labels, proposed recommendation methods, and identified mentoring as a key success factor, longitudinal contribution dynamics in social-good OSS have been examined~\cite{Fang2026}, but GFI-specific labeling practices and newcomer PR outcomes in mainstream repositories remain unexplored. Our study takes a longitudinal perspective, examining these trends over four years across 37 popular OSS projects.

\begin{figure*}[t]
\centering
\includegraphics[width=\linewidth]{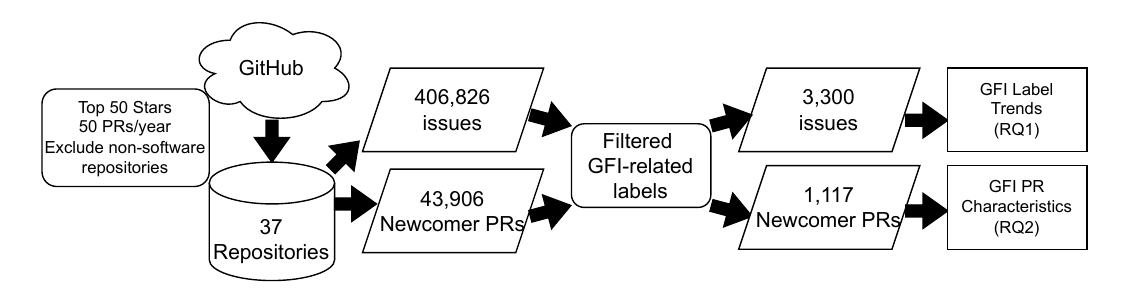}
\caption{Overview of the study method.}
\Description{Overview of the study methodology showing repository selection, data collection of GFI issues and newcomer pull requests, and analyses for RQ1 (GFI trends and newcomer engagement) and RQ2 (PR characteristics and merge rate trends).}
\label{fig:overview}
\end{figure*}

\section{Methodology}
\label{sec:methodology}
Figure~\ref{fig:overview} illustrates the overview of our research methodology. We selected our target repositories from the top 50 most-starred repositories on GitHub with at least 50 pull requests per year, collected issues and newcomer pull requests, and analyzed GFI label trends and newcomer engagement (RQ1) and characteristics of GFI PRs and merge rates over time (RQ2).

\subsection{Repository Selection}

Repository selection proceeded in three steps. First, to ensure sustained contribution activity throughout the study period, we applied a minimum activity threshold: only repositories that received at least 50 pull requests in each of the four calendar years from 2021 to 2024 were considered as candidates. This per-year requirement avoids including repositories that are sporadically active or became inactive during the study period. Second, the qualifying repositories were ranked by total star count, and the top 50 most-starred were retained as the initial set. Third, we excluded repositories that do not constitute software projects (e.g., tutorial collections, curated lists, and learning resources) and those that had disabled GitHub Issues, as issue tracking is a prerequisite for GFI-based workflows. This three-step process yielded a final set of 37 software repositories for analysis.

The selected repositories are diverse, with the distribution of primary programming languages being TypeScript (9 repositories, 24.3\%), C++ (5, 13.5\%), JavaScript (5, 13.5\%), Python (5, 13.5\%), Rust (3, 8.1\%), and others (10, 27.0\%).

\subsection{Data Collection}

\subsubsection{Issue Data Collection (RQ1)}

To answer RQ1, we used the GitHub GraphQL API to collect all GFI-labeled issues from July 2021 to June 2025. To identify GFIs, we followed the method adopted by Turzo et al.~\cite{10.1109/TSE.2025.3550881}, combining the label list presented by Tan et al.~\cite{10.1145/3368089.3409746} with newcomer contribution guidelines~\cite{gazanchyan2020awesome}. For each issue, we recorded its creation date, label information, and closed state.

We collected 406,826 issues, of which 3,300 were GFI-labeled, to compute the monthly GFI ratio (proportion of GFI-labeled issues per month). Pull requests addressing these GFI issues were also collected to analyze newcomer engagement.

\subsubsection{Pull Request Data Collection (RQ2)}

To answer RQ2, we collected pull requests that address GFI-labeled issues (GFI PRs) using the GitHub GraphQL API. A PR was considered to address a GFI-labeled issue if its body contained a GitHub closing keyword followed by an issue number. We define a \textit{newcomer} as a contributor submitting their \textit{first-ever pull request to a specific repository} (not their first contribution to OSS in general). Newcomers were therefore identified on a per-repository basis: we extracted the first-ever pull request submitted by each user to a given repository between July 2021 and June 2025. For each pull request, we retrieved the PR number, title, body, creation date, merge date, state (MERGED, CLOSED, OPEN), lines added, lines deleted, number of changed files, commit count, review comment count, and label information. For the description, we measured the length of substantive user-written content after removing HTML comments from PR templates. To ensure data quality, we filtered out bots and deleted accounts using the author\textunderscore type field from the GitHub API, retaining only those where author\textunderscore type was `User'. All data was collected via the GitHub API in November 2025, approximately five months after the end of the study period. This observation buffer exceeds the 95th percentile of time-to-merge among merged PRs (approximately 80 days), ensuring that PRs created near the end of the analysis window had sufficient time to be reviewed and resolved. For merge rate calculations, only MERGED PRs were counted as merged; CLOSED and OPEN PRs (including 48 still-open) were treated as unmerged. For insertions and deletions, which exhibited highly skewed distributions, we applied a log transformation. This process yielded 1,117 newcomer GFI PRs, which serve as the dataset for merge rate analyses.

For time-series comparison, we divided the four-year period into 12-month analysis years: Y1 (Jul 2021--Jun 2022), Y2 (Jul 2022--Jun 2023), Y3 (Jul 2023--Jun 2024), and Y4 (Jul 2024--Jun 2025). We classified each PR into one of four task types (Bug, Feature, Documentation, or Other) based on the labels of its referenced GFI issue. A label was mapped to Bug if it contained ``bug'' as a word boundary, to Feature if it contained ``feature'' or ``enhancement'', to Documentation if it contained ``doc'', and to Other otherwise. Word-boundary matching for ``bug'' prevents false positives from area labels such as ``debug.'' For task-type analyses, we further classified each PR based on its referenced GFI issue's labels; of the 1,117 PRs, 1,070 matched exactly one task type and were included, while the remaining 47 matching multiple types were excluded to avoid ambiguous classification.

To control for multiple comparisons, we applied Holm-Bonferroni correction within each analysis table, and Benjamini-Hochberg correction ($\alpha=0.05$) for the 30 repository-level trend tests.

\section{Results}
\label{sec:results}
\subsection{RQ1: How have GFI practices and newcomer engagement changed over time?}

\subsubsection{GFI Ratio Trend}

\begin{figure}[t]
\centering
\includegraphics[width=\columnwidth]{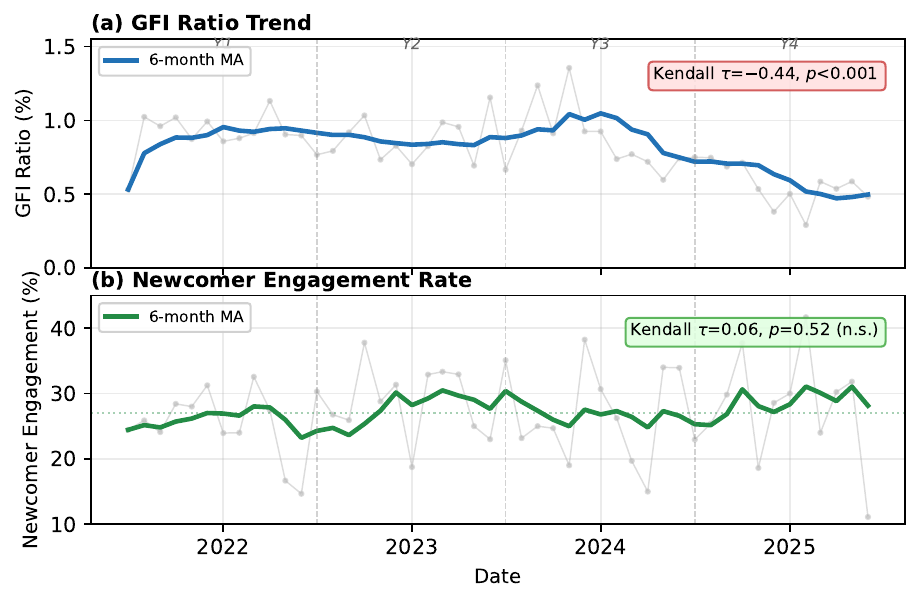}
\caption{RQ1 time-series trends. (a) Monthly GFI ratio shows a significant decreasing trend ($\tau=-0.44$, $p<0.001$). (b) Newcomer engagement rate remains stable at approximately 27\% ($\tau=0.06$, n.s.). Gray dots: monthly values; colored lines: 6-month moving averages; dashed vertical lines: analysis year boundaries.}
\Description{Two time-series plots for RQ1. Left: monthly GFI ratio (percentage of issues labeled good-first-issue) from July 2021 to June 2025, showing a significant decreasing trend from around 0.9\% to 0.6\%. Right: monthly newcomer engagement rate showing stable values around 27\% throughout the period with no significant trend.}
\label{fig:rq1_trends}
\end{figure}

A Mann-Kendall trend test on the monthly GFI ratio confirmed a statistically significant decreasing trend ($\tau=-0.44$, $p<0.001$; Table~\ref{tab:rq1_trend}), though as shown in Figure~\ref{fig:rq1_trends}(a), the decline was not gradual: the yearly average remained stable from Y1 (0.92\%) through Y3 (0.88\%) before dropping sharply in Y4 (0.57\%).

A Pettitt change-point test identified a structural break at January 2024 ($K=445$, $p<0.001$): the mean GFI ratio was 0.91\% before and 0.63\% after, a 31\% decrease.

To verify that the Y4 decline is not a data-collection artifact caused by labeling lag, we performed two checks. First, repositories with reduced GFI counts in Y4 continued creating issues at their usual volume yet assigned zero GFI labels for seven to nine months, a duration far exceeding plausible triage delays. Second, a GitHub Timeline API verification across all 3,300 GFI issues confirmed that 90\% received their label within 67 days of issue creation (90th percentile). An adjusted GFI ratio accounting for this empirical capture rate confirms the decline (Y4: 0.59\% vs.\ Y3: 0.89\%), ruling out a labeling-lag explanation.

\subsubsection{Repository Heterogeneity in GFI Trends}

Of the 37 repositories, 7 (18.9\%) never used GFI labels and were excluded from trend analysis. Among the remaining 30, Mann-Kendall tests (BH-corrected) revealed substantial heterogeneity: 7 (23.3\%) showed a decreasing trend, 21 (70.0\%) no significant trend, and 2 (6.7\%) an increasing trend. This variation was not explained by repository age (Spearman's $\rho=-0.105$, $p=0.588$) or primary programming language.

Comparing characteristics across the three trend groups (Kruskal-Wallis test), no statistically significant differences were found in star count, repository age, total issue count, or GFI count (all $p>0.2$). These results suggest that GFI usage trends cannot be explained by objective characteristics such as project size, maturity, or primary language, but rather depend heavily on project-specific strategic decisions. To confirm that the aggregate decline is not dominated by a few projects, we re-ran the Mann-Kendall test after excluding the five repositories with the largest ratio declines; the result remained significant ($\tau=-0.275$, $p=0.006$), indicating that the decline is broadly distributed across repositories.

\begin{table}[t]
\centering
\caption{Time-series trend analysis of RQ1 metrics (Mann-Kendall test)}
\label{tab:rq1_trend}
\footnotesize
\setlength{\tabcolsep}{3pt}
\begin{tabular}{lrrrrcc}
\toprule
\textbf{Metric} & \textbf{Y1} & \textbf{Y2} & \textbf{Y3} & \textbf{Y4} & \textbf{Kendall $\tau$} & \textbf{Trend} \\
\midrule
GFI Ratio (\%)           & 0.92 & 0.87 & 0.88 & 0.57 & -0.44*** & Decr. \\
Newcomer Engagement (\%) & 25.1 & 28.9 & 27.1 & 27.7 & 0.06     & --    \\
\bottomrule
\end{tabular}
\vspace{1mm}
\parbox{\linewidth}{\footnotesize Note: **$p<0.01$, ***$p<0.001$ (Holm-Bonferroni adjusted). Monthly Mann-Kendall over 48 months; Y1--Y4: yearly averages.}
\end{table}

\begin{table}[t]
\centering
\caption{Merge rate by task type and analysis year}
\label{tab:merge_by_type}
\small
\setlength{\tabcolsep}{3pt}
\begin{tabular}{lrrrrrc}
\toprule
\textbf{Task Type} & \textbf{Y1} & \textbf{Y2} & \textbf{Y3} & \textbf{Y4} & \textbf{Total} & \textbf{Trend} \\
\midrule
Bug   & 64.5\% & 83.5\% & 71.9\% & 45.9\% & 68.7\% & Decr.* \\
Feature & 53.7\% & 54.8\% & 53.3\% & 55.6\% & 54.4\% & None \\
Docs  & 68.4\% & 65.2\% & 42.4\% & 47.7\% & 52.9\% & Decr.* \\
Other & 57.1\% & 46.4\% & 33.3\% & 28.6\% & 40.7\% & Decr.* \\
\bottomrule
\end{tabular}
\vspace{1mm}
\parbox{\linewidth}{\footnotesize Note: *$p<0.05$, **$p<0.01$, ***$p<0.001$ (Mann-Kendall test, Holm-Bonferroni adjusted). Cell $n$ (Y1/Y2/Y3/Y4): Bug=97/96/85/53; Feature=40/41/28/35; Docs=19/22/32/40; Other=98/138/120/126.}
\end{table}

\subsubsection{Newcomer Engagement with GFIs}

As shown in Figure~\ref{fig:rq1_trends}(b), the proportion of GFI issues addressed by newcomers remained stable throughout the period. The overall engagement rate was 27.0\% (891 out of 3,300 GFI issues), with per-year rates between 25\% and 29\%. A Mann-Kendall trend test confirmed no significant trend ($\tau=0.06$, $p=0.52$). Despite the declining GFI ratio, the proportion of GFI issues addressed by newcomers showed no significant change.

\begin{acmbox}{RQ1 Summary}
Three key findings emerge: (1) The GFI ratio remained stable for three years before declining significantly after January 2024, dropping from 0.91\% to 0.63\%. (2) This trend varied substantially across repositories (23.3\% decreasing, 70.0\% stable, 6.7\% increasing), driven by project-specific decisions rather than observable characteristics. (3) Newcomer engagement remained stable at 27.0\%, indicating that demand for GFI-based contribution opportunities persists despite declining supply.
\end{acmbox}

\subsection{RQ2: How have the characteristics and task types of newcomer GFI PRs changed over time, and are these factors associated with merge rates?}

\subsubsection{PR Metrics Trends and Task Type Analysis}
\label{sec:results_rq2}

The 1,117 GFI PRs spanned all 30 GFI-using repositories, with no single repository exceeding 17\% of the total. The overall merge rate was 53.0\%. As shown in Table~\ref{tab:rq2_trend}, the merge rate remained stable through Y2 (61.9\%) before declining to 42.2\% in Y4 ($\tau=-0.35$, $p<0.01$), while description length increased sharply from Y1 (306) to Y2 (439) before plateauing around 480--490 through Y3 and Y4 ($\tau=0.35$, $p<0.01$).

\begin{table}[t]
\centering
\caption{Time-series trend analysis of RQ2 metrics (Mann-Kendall test)}
\label{tab:rq2_trend}
\footnotesize
\setlength{\tabcolsep}{3pt}
\begin{tabular}{lrrrrcc}
\toprule
\textbf{Metric} & \textbf{Y1} & \textbf{Y2} & \textbf{Y3} & \textbf{Y4} & \textbf{Kendall $\tau$} & \textbf{Trend} \\
\midrule
Merge Rate (\%)    & 61.9 & 61.9 & 48.5 & 42.2 & -0.35** & Decr. \\
Description Length & 306  & 439  & 486  & 481  & 0.35**  & Incr. \\
\bottomrule
\end{tabular}
\vspace{1mm}
\parbox{\linewidth}{\footnotesize Note: **$p<0.01$, ***$p<0.001$ (Holm-Bonferroni adjusted). Monthly Mann-Kendall over 48 months; Y1--Y4: yearly averages.}
\end{table}

As shown in Table~\ref{tab:merge_by_type}, bug-fix tasks had the highest overall merge rate (68.7\%), while Feature tasks remained stable at approximately 54\% with no significant trend.

\begin{table}[t]
\centering
\caption{GFI PR metrics by merge status}
\label{tab:merge_factors}
\small
\setlength{\tabcolsep}{3pt}
\begin{tabular}{lrrcc}
\toprule
\textbf{Metric} & \textbf{Merged} & \textbf{Not Merged} & \textbf{p-value} & \textbf{$|r|$} \\
\midrule
\multicolumn{5}{l}{\textit{Initial PR characteristics}} \\
Insertions (log)    & 3.02  & 2.89  & 0.885   & 0.01 \\
Deletions (log)     & 1.10  & 1.39  & 0.994   & 0.00 \\
Changed Files       & 2.0   & 2.0   & 0.155   & 0.05 \\
Description Length  & 382.5 & 435.0 & 0.608   & 0.02 \\
\midrule
\multicolumn{5}{l}{\textit{Process-level metrics\textsuperscript{\dag}}} \\
Commits Count       & 3.0   & 2.0   & $<$0.01**   & 0.15 \\
Review Count        & 2.0   & 1.0   & $<$0.001*** & 0.32 \\
\bottomrule
\end{tabular}
\vspace{1mm}
\parbox{\linewidth}{\footnotesize Note: **$p<0.01$, ***$p<0.001$ (Holm-Bonferroni adjusted). Median values shown. Mann-Whitney U test. $|r|$: rank-biserial effect size. \textsuperscript{\dag}Process-level metrics accumulate during the review lifecycle and are subject to mechanical confounds (see text).}
\end{table}

\subsubsection{Factors Associated with Merge Success}

As shown in Table~\ref{tab:merge_factors}, none of the initial PR characteristics (code size, changed files, description length) were significantly associated with merge outcomes. Although commit count and review count differed significantly, both accumulate through the review lifecycle and cannot be interpreted as independent predictors of merge outcomes.

\begin{acmbox}{RQ2 Summary}
Two key findings emerge: (1) Newcomer GFI PRs evolved over time, with description length increasing significantly while merge rates remained stable through Y2 before declining from 61.9\% to 42.2\%. Bug-fix tasks showed the highest but declining merge rate (68.7\%) and Feature tasks remained stable (54.4\%). (2) Initial PR characteristics showed no significant association with merge outcomes, indicating that merge success is not predicted by the quantitative characteristics of the initial submission.
\end{acmbox}

\section{Discussion}
\label{sec:discussion}
\subsection{Interpretation of RQ1 Findings}

The decline in GFI ratio was not gradual: as shown in Table~\ref{tab:rq1_trend}, values remained stable through Y3 before dropping sharply in Y4. The Pettitt change-point test locates a structural break at January 2024, indicating that the shift began in the second half of Y3 and accelerated through Y4. This trend varied substantially across repositories and could not be explained by repository age or programming language.

These results suggest that changes in GFI usage are not uniformly driven by a single external factor, but are strongly dependent on project-specific strategic decisions. A qualitative inspection of the seven decreasing-trend repositories' CONTRIBUTING.md and issue template histories reveals heterogeneous causes. For instance, yt-dlp/yt-dlp introduced anti-spam protections in August 2024, after which GFI labeling dropped sharply, and godotengine/godot assigned zero GFI labels for 17 consecutive months while updating its contribution guidelines to mitigate spam mid-gap, with labeling only resuming afterward. Notably, both cases coincide with the period of rapid generative AI tool adoption (post-2023), suggesting that an influx of low-quality or AI-generated contributions may have eroded the attractiveness of GFI labels for maintainers, though this remains anecdotal and systematic measurement is a direction for future work. In contrast, other repositories such as vercel/next.js showed no discernible documentation changes accompanying their decline, suggesting that informal shifts in maintainer practice also play a role. Conversely, projects actively seeking to attract new contributors may be increasing their use of GFIs.

The proportion of GFIs addressed by newcomers remained stable at approximately 27\% throughout the four-year period, with no significant trend. Despite the GFI ratio decline, newcomers continued to engage with available GFI issues at a consistent rate. However, 73\% of GFIs did not receive a newcomer PR, consistent with the observation by Tan et al.~\cite{10.1145/3368089.3409746} that many GFIs are not taken up by newcomers.

\subsection{Interpretation of RQ2 Findings}
\label{sec:discussion_rq2}

Merge rates declined across most task types, with the notable exception of Feature tasks, which remained stable throughout the period (Table~\ref{tab:merge_by_type}). This non-uniformity indicates that the decline is not driven by a single global factor.

Description length increased significantly over the study period (+57\%). Despite submitting longer descriptions, unmerged PRs did not differ significantly from merged PRs in this metric (Table~\ref{tab:merge_factors}). Multiple factors may contribute to this trend. A Pettitt change-point test on the monthly median description length series identifies a structural break at January 2023 ($K=332$, $p=0.006$), coinciding with the rapid adoption of ChatGPT. While the yearly average for Y2 (Jul 2022--Jun 2023) is higher than Y1, the monthly analysis reveals that the increase concentrated in the second half of Y2, consistent with AI-assisted writing tool adoption rather than predating it.

Process-level metrics showed statistically significant differences, but should not be interpreted as independent predictors. Review count in particular is prone to structural confounds: branch protection rules and selective maintainer attention may both inflate review counts for merged PRs irrespective of PR quality. We therefore do not treat these metrics as predictors of merge success.

The parallel declines in GFI supply (RQ1) and merge rate (RQ2) suggest that the challenges to newcomer onboarding extend beyond task availability.

\subsection{Implications for Practice}

\textbf{For Project Maintainers}: Our findings indicate that GFI labels continue to attract a stable proportion of newcomers (approximately 27\%). As both the GFI ratio and newcomer PR merge rate show declining trends, actively creating and labeling GFIs while sustaining review support for newcomer PRs may help maintain the effectiveness of GFI-based onboarding. Regardless of the underlying cause, the stable newcomer engagement rate suggests that demand for GFI-based contribution opportunities remains high.

\textbf{For Newcomers}: A key finding is that despite newcomers submitting increasingly detailed PRs over the study period (description length increased by 57\%), merge rates continued to decline. The quantitative characteristics of the initial submission do not predict merge outcomes, leaving open the question of what does. Prior work has found that maintainer mentoring, not initial submission quality, is the primary driver of newcomer success~\cite{10.1109/ICSE48619.2023.00064}; newcomers should therefore seek out projects and maintainers that actively support contributions, and proactively engage with reviewer feedback once a PR is submitted.

\textbf{For Researchers}: This study emphasizes the importance of considering project heterogeneity in OSS onboarding research. Analyses based solely on aggregated statistics may overlook the diverse strategies adopted by individual projects. Additionally, our label-based task-type analysis demonstrates that more granular categorization can yield actionable insights. However, researchers should be aware that labeling practices vary across projects and should be accounted for in cross-project studies.

\section{Threats to Validity}
\label{sec:threats}
\textbf{Construct Validity}: We used the GFI label list from Tan et al.~\cite{10.1145/3368089.3409746} and Turzo et al.~\cite{10.1109/TSE.2025.3550881}; projects with custom naming conventions may not be fully captured. Our task-type classification (Bug, Feature, Documentation, Other) is keyword-based; the 47 cases (4.2\%) matching multiple types were excluded to avoid ambiguous classification. We define newcomers as first-time PR submitters within a repository, which does not account for prior GitHub experience. Consequently, experienced developers joining a new repository are classified alongside true novices. Our ``newcomers'' therefore represent repository newcomers rather than OSS novices; if restricted to contributors with no prior OSS activity, merge rates and PR characteristics might differ. GFI PRs were identified via closing keywords in PR bodies; informal references (e.g., ``related to \#123'') may not have been captured. Bot detection relies on GitHub's \texttt{author\_type} field, which may not cover all automated contributions. Of the 48 still-open PRs, 27 (56.3\%) fall in Y4, 10 (20.8\%) in Y3, 7 (14.6\%) in Y2, and 4 (8.3\%) in Y1; if eventually merged, the Y4 rate could reach 52.8\%, though our five-month observation buffer exceeds the 95th-percentile time-to-merge ($\approx$80 days).

\textbf{Internal Validity}: This study observes time-series trends and does not claim causal relationships; observed changes may reflect multiple confounding factors, including ecosystem-wide shifts, project-specific decisions, and evolving development tools. In RQ2, 95 GFI issues (9.7\%) received multiple newcomer PRs; treating each as an independent observation may introduce non-independence. Furthermore, the study period coincides with the rapid adoption of generative AI tools for software contribution~\cite{10.1145/3643773}. AI-generated contributions may superficially satisfy formal criteria (e.g., referencing a GFI issue, containing code changes) while exhibiting quality issues that reduce their acceptability~\cite{yetistiren2023evaluating}, and detecting such contributions from quantitative signals alone remains an open challenge~\cite{10.1109/ICSE55347.2025.00064}. Such submissions may therefore have contributed to the observed decline in merge rates.

\textbf{External Validity}: This study is limited to popular OSS projects (top-starred) on GitHub. Different trends may be observed in smaller projects or on other platforms (e.g., GitLab, Bitbucket). Furthermore, the analysis is confined to software projects, excluding non-software repositories such as documentation and learning resources.

\section{Conclusion}
\label{sec:conclusion}
From 406,826 issues across 37 popular OSS projects on GitHub, we identified 3,300 GFI-labeled issues and analyzed 1,117 newcomer GFI PRs to investigate the GFI ratio, newcomer engagement patterns, and the changing characteristics of GFI PRs over a four-year period from July 2021 to June 2025.

Regarding RQ1, a change-point analysis identified a structural break in January 2024, after which the GFI ratio declined significantly. This trend varied greatly among repositories, driven by project-specific decisions rather than observable project characteristics. Newcomer engagement remained stable at approximately 27\%, indicating sustained demand despite declining supply.

Regarding RQ2, the merge rate declined over the period. Bug-fix tasks maintained the highest merge rate (68.7\%), and Feature tasks showed no significant trend. Initial PR characteristics showed no association with merge outcomes, suggesting that GFIs remain appropriately scoped.

Together, these findings reveal a widening gap between stable newcomer interest in GFIs and the declining availability and success of GFI-based onboarding, underscoring the need for maintainers to sustain both GFI labeling and review support.

Future research would benefit from (1) a qualitative investigation into the decision-making processes behind why some projects are increasing their use of GFIs while others are decreasing it, and (2) identification of the root causes for the declining GFI PR merge rates (e.g., issue quality, maintenance resources, or changes in quality standards). These open questions also motivate the development of AI-assisted tools for automatically detecting and recommending GFI candidates, which could reduce the labeling burden on maintainers and help counteract the observed decline in GFI availability.

\section*{Data Availability}
The replication package for this study is available at \url{https://doi.org/10.5281/zenodo.18847761}.

\section*{Acknowledgement}
This research has been supported by JSPS KAKENHI Nos. JP24K14895 and JP26K21197, and JST BOOST, Japan Grant Number JPMJBS2423.

\bibliographystyle{ACM-Reference-Format}
\bibliography{reference}

@ARTICLE{9273270,
  author={Subramanian, Vikram N. and Rehman, Ifraz and Nagappan, Meiyappan and Kula, Raula Gaikovina},
  journal={IEEE Software},
  title={Analyzing First Contributions on {GitHub}: What Do Newcomers Do?},
  year={2022},
  volume={39},
  number={1},
  pages={93--101},
  doi={10.1109/MS.2020.3041241}}

@article{10.1109/TSE.2025.3550881,
author = {Turzo, Asif Kamal and Sultana, Sayma and Bosu, Amiangshu},
title = {From First Patch to Long-Term Contributor: Evaluating Onboarding Recommendations for {OSS} Newcomers},
year = {2025},
issue_date = {April 2025},
publisher = {IEEE Press},
volume = {51},
number = {4},
issn = {0098-5589},
journal = {IEEE Trans. Softw. Eng.},
month = apr,
pages = {1303--1318},
numpages = {16}
}

@inproceedings{10.1145/3368089.3409746,
author = {Tan, Xin and Zhou, Minghui and Sun, Zeyu},
title = {A first look at good first issues on {GitHub}},
year = {2020},
isbn = {9781450370431},
publisher = {Association for Computing Machinery},
booktitle = {Proceedings of the 28th ACM Joint Meeting on European Software Engineering Conference and Symposium on the Foundations of Software Engineering},
pages = {398--409},
numpages = {12},
location = {Virtual Event, USA}
}

@inproceedings{10.1145/3510003.3510196,
author = {Xiao, Wenxin and He, Hao and Xu, Weiwei and Tan, Xin and Dong, Jinhao and Zhou, Minghui},
title = {Recommending good first issues in {GitHub} {OSS} projects},
year = {2022},
isbn = {9781450392211},
publisher = {Association for Computing Machinery},
booktitle = {Proceedings of the 44th International Conference on Software Engineering},
pages = {1830--1842},
numpages = {13},
location = {Pittsburgh, Pennsylvania}
}

@inproceedings{10.1145/3475716.3475789,
author = {Huang, Yuekai and Wang, Junjie and Wang, Song and Liu, Zhe and Wang, Dandan and Wang, Qing},
title = {Characterizing and Predicting Good First Issues},
year = {2021},
isbn = {9781450386654},
publisher = {Association for Computing Machinery},
booktitle = {Proceedings of the 15th ACM / IEEE International Symposium on Empirical Software Engineering and Measurement (ESEM)},
articleno = {13},
numpages = {12},
location = {Bari, Italy}
}

@inproceedings{10.1145/3544902.3546236,
author = {Santos, Fabio and Trinkenreich, Bianca and Pimentel, Jo\~{a}o Felipe and Wiese, Igor and Steinmacher, Igor and Sarma, Anita and Gerosa, Marco A.},
title = {How to Choose a Task? Mismatches in Perspectives of Newcomers and Existing Contributors},
year = {2022},
isbn = {9781450394277},
publisher = {Association for Computing Machinery},
booktitle = {Proceedings of the 16th ACM / IEEE International Symposium on Empirical Software Engineering and Measurement},
pages = {114--124},
numpages = {11},
location = {Helsinki, Finland}
}

@inproceedings{10.1145/2884781.2884806,
author = {Steinmacher, Igor and Conte, Tayana Uchoa and Treude, Christoph and Gerosa, Marco Aur\'{e}lio},
title = {Overcoming open source project entry barriers with a portal for newcomers},
year = {2016},
isbn = {9781450339001},
publisher = {Association for Computing Machinery},
booktitle = {Proceedings of the 38th International Conference on Software Engineering},
pages = {273--284},
numpages = {12},
location = {Austin, Texas}
}

@INPROCEEDINGS{6943482,
  author={Steinmacher, Igor and Chaves, Ana Paula and Conte, Tayana Uchoa and Gerosa, Marco Aurelio},
  booktitle={2014 Brazilian Symposium on Software Engineering},
  title={Preliminary Empirical Identification of Barriers Faced by Newcomers to Open Source Software Projects},
  year={2014},
  volume={},
  number={},
  pages={51--60},
  doi={10.1109/SBES.2014.9}}

@inproceedings{10.1145/2675133.2675215,
author = {Steinmacher, Igor and Conte, Tayana and Gerosa, Marco Aur\'{e}lio and Redmiles, David},
title = {Social Barriers Faced by Newcomers Placing Their First Contribution in Open Source Software Projects},
year = {2015},
isbn = {9781450329224},
publisher = {Association for Computing Machinery},
booktitle = {Proceedings of the 18th ACM Conference on Computer Supported Cooperative Work \& Social Computing},
pages = {1379--1392},
numpages = {14},
location = {Vancouver, BC, Canada}
}

@Article{Steinmacher2021,
author={Steinmacher, Igor
and Balali, Sogol
and Trinkenreich, Bianca
and Guizani, Mariam
and Izquierdo-Cortazar, Daniel
and Cuevas Zambrano, Griselda G.
and Gerosa, Marco Aurelio
and Sarma, Anita},
title={Being a Mentor in open source projects},
journal={Journal of Internet Services and Applications},
year={2021},
month={Sep},
day={09},
volume={12},
number={1},
pages={7},
issn={1869-0238},
doi={10.1186/s13174-021-00140-z},
url={https://doi.org/10.1186/s13174-021-00140-z}
}

@inproceedings{10.1145/3412569.3412571,
author = {Balali, Sogol and Annamalai, Umayal and Padala, Hema Susmita and Trinkenreich, Bianca and Gerosa, Marco A. and Steinmacher, Igor and Sarma, Anita},
title = {Recommending Tasks to Newcomers in {OSS} Projects: How Do Mentors Handle It?},
year = {2020},
isbn = {9781450387798},
publisher = {Association for Computing Machinery},
booktitle = {Proceedings of the 16th International Symposium on Open Collaboration},
articleno = {7},
numpages = {14},
location = {Virtual conference, Spain}
}

@inproceedings{10.1109/ICSE48619.2023.00064,
author = {Tan, Xin and Chen, Yiran and Wu, Haohua and Zhou, Minghui and Zhang, Li},
title = {Is it Enough to Recommend Tasks to Newcomers? Understanding Mentoring on Good First Issues},
year = {2023},
isbn = {9781665457019},
publisher = {IEEE Press},
booktitle = {Proceedings of the 45th International Conference on Software Engineering},
pages = {653--664},
numpages = {12},
location = {Melbourne, Victoria, Australia}
}

@article{10.1145/3643773,
author = {Xiao, Tao and Hata, Hideaki and Treude, Christoph and Matsumoto, Kenichi},
title = {Generative {AI} for Pull Request Descriptions: Adoption, Impact, and Developer Interventions},
year = {2024},
issue_date = {July 2024},
publisher = {Association for Computing Machinery},
volume = {1},
number = {FSE},
journal = {Proc. ACM Softw. Eng.},
month = jul,
articleno = {47},
numpages = {23}
}

@article{yetistiren2023evaluating,
  author       = {Yetiştiren, Burak and Özsoy, Işık and Ayerdem, Miray and Tüzün, Eray},
  title        = {Evaluating the Code Quality of {AI}-Assisted Code Generation Tools: An Empirical Study on {GitHub} {Copilot}, {Amazon} {CodeWhisperer}, and {ChatGPT}},
  year         = {2023},
  month        = {Apr},
  journal      = {arXiv preprint arXiv:2304.10778},
  archivePrefix = {arXiv},
  eprint       = {2304.10778}
}

@inproceedings{10.1109/ICSE55347.2025.00064,
author = {Suh, Hyunjae and Tafreshipour, Mahan and Li, Jiawei and Bhattiprolu, Adithya and Ahmed, Iftekhar},
title = {An Empirical Study on Automatically Detecting {AI}-Generated Source Code: How Far Are We?},
year = {2025},
isbn = {9798331505691},
publisher = {IEEE Press},
booktitle = {Proceedings of the IEEE/ACM 47th International Conference on Software Engineering},
pages = {859--871},
numpages = {13},
location = {Ottawa, Ontario, Canada}
}

@inproceedings{10.1145/3510458.3513020,
author = {Guizani, Mariam and Zimmermann, Thomas and Sarma, Anita and Ford, Denae},
title = {Attracting and retaining {OSS} contributors with a maintainer dashboard},
year = {2022},
isbn = {9781450392273},
publisher = {Association for Computing Machinery},
booktitle = {Proceedings of the 2022 ACM/IEEE 44th International Conference on Software Engineering: Software Engineering in Society},
pages = {36--40},
numpages = {5},
location = {Pittsburgh, Pennsylvania}
}

@inproceedings{10.1109/ASE56229.2023.00158,
author = {Xiao, Wenxin and Li, Jingyue and He, Hao and Qiu, Ruiqiao and Zhou, Minghui},
title = {Personalized First Issue Recommender for Newcomers in Open Source Projects},
year = {2024},
isbn = {9798350329964},
publisher = {IEEE Press},
booktitle = {Proceedings of the 38th IEEE/ACM International Conference on Automated Software Engineering},
pages = {800--812},
numpages = {13},
location = {Echternach, Luxembourg}
}

@article{sholler2019ten,
  title={Ten simple rules for helping newcomers become contributors to open projects},
  author={Sholler, Dan and Steinmacher, Igor and Ford, Denae and Averick, Mara and Hoye, Mike and Wilson, Greg},
  journal={PLoS computational biology},
  volume={15},
  number={9},
  pages={e1007296},
  year={2019},
  publisher={Public Library of Science San Francisco, CA USA},
  doi={10.1371/journal.pcbi.1007296}
}

@misc{gazanchyan2020awesome,
  author = {Gazanchyan, Sergey},
  title  = {Awesome first {PR} opportunities},
  year   = {2020},
  note   = {[Online]. Available: \url{https://github.com/MunGell/awesome-for-beginners}}
}

@article{Fang2026,
  author    = {Fang, Zihan and Zhang, Yueke and Zimmermann, Thomas and Ford, Denae and Huang, Yu},
  title     = {Contribution Patterns in Open Source Software for Social Good: Dynamics, Individuals, and Impact},
  journal   = {Proceedings of the ACM on Human-Computer Interaction},
  volume    = {10},
  number    = {CSCW},
  article   = {CSCW010},
  year      = {2026},
  month     = apr,
  numpages  = {26},
  doi       = {10.1145/3788046},
  publisher = {ACM}
}

\end{document}